\documentstyle[epsf]{mn}

\newif\ifAMStwofonts

\ifoldfss
  \ifCUPmtlplainloaded \else
    \NewTextAlphabet{textbfit} {cmbxti10} {}
    \NewTextAlphabet{textbfss} {cmssbx10} {}
    \NewMathAlphabet{mathbfit} {cmbxti10} {} 
    \NewMathAlphabet{mathbfss} {cmssbx10} {} 
  \fi
  \ifAMStwofonts
    \ifCUPmtlplainloaded \else
      \NewSymbolFont{upmath} {eurm10}
      \NewSymbolFont{AMSa} {msam10}
      \NewMathSymbol{\upi}     {0}{upmath}{19}
      \NewMathSymbol{\umu}     {0}{upmath}{16}
      \NewMathSymbol{\upartial}{0}{upmath}{40}
      \NewMathSymbol{\leqslant}{3}{AMSa}{36}
      \NewMathSymbol{\geqslant}{3}{AMSa}{3E}

    \fi
  \fi
\fi 

\ifnfssone
  \newmathalphabet{\mathit}
  \addtoversion{normal}{\mathit}{cmr}{m}{it}
  \addtoversion{bold}{\mathit}{cmr}{bx}{it}
  \newmathalphabet{\mathbfit} 
  \addtoversion{normal}{\mathbfit}{cmr}{bx}{it}
  \addtoversion{bold}{\mathbfit}{cmr}{bx}{it}
  \newmathalphabet{\mathbfss} 
  \addtoversion{normal}{\mathbfss}{cmss}{bx}{n}
  \addtoversion{bold}{\mathbfss}{cmss}{bx}{n}
  \ifAMStwofonts
    \ifCUPmtlplainloaded \else
      \UseAMStwoboldmath
      \makeatletter
      \new@mathgroup\upmath@group
      \define@mathgroup\mv@normal\upmath@group{eur}{m}{n}
      \define@mathgroup\mv@bold\upmath@group{eur}{b}{n}
      \edef\UPM{\hexnumber\upmath@group}
      \new@mathgroup\amsa@group
      \define@mathgroup\mv@normal\amsa@group{msa}{m}{n}
      \define@mathgroup\mv@bold\amsa@group{msa}{m}{n}
      \edef\AMSa{\hexnumber\amsa@group}
      \makeatother
      \mathchardef\upi="0\UPM19
      \mathchardef\umu="0\UPM16
      \mathchardef\upartial="0\UPM40
      \mathchardef\leqslant="3\AMSa36
      \mathchardef\geqslant="3\AMSa3E
    \fi
  \fi
\fi 

\ifnfsstwo
  \DeclareMathAlphabet{\mathbfit}{OT1}{cmr}{bx}{it}
  \SetMathAlphabet\mathbfit{bold}{OT1}{cmr}{bx}{it}
  \DeclareMathAlphabet{\mathbfss}{OT1}{cmss}{bx}{n}
  \SetMathAlphabet\mathbfss{bold}{OT1}{cmss}{bx}{n}
  \ifAMStwofonts
    \ifCUPmtlplainloaded \else
      \DeclareSymbolFont{UPM}{U}{eur}{m}{n}
      \SetSymbolFont{UPM}{bold}{U}{eur}{b}{n}
      \DeclareSymbolFont{AMSa}{U}{msa}{m}{n}
      \DeclareMathSymbol{\upi}{0}{UPM}{"19}
      \DeclareMathSymbol{\umu}{0}{UPM}{"16}
      \DeclareMathSymbol{\upartial}{0}{UPM}{"40}
      \DeclareMathSymbol{\leqslant}{3}{AMSa}{"36}
      \DeclareMathSymbol{\geqslant}{3}{AMSa}{"3E}
    \fi
  \fi
\fi 

\ifCUPmtlplainloaded \else
  \ifAMStwofonts \else 
    \def\upi{\pi}
    \def\umu{\mu}
    \def\upartial{\partial}
  \fi
\fi

\title{Density waves in the shearing sheet \\
       III. Disc heating}
\author[B. Fuchs]
       {B. Fuchs \\
        Astronomisches Rechen--Institut, M\"onchhofstr.~12--14, 69120
Heidelberg, Germany}
\date{Accepted
      Received ;
      in original form 2000 May}

\pagerange{\pageref{firstpage}--\pageref{lastpage}}
\pubyear{2001}

\begin{document}

\maketitle

\label{firstpage}

\begin{abstract}
The problem of dynamical heating of galactic discs by spiral density waves is
discussed using the shearing sheet model. The secular evolution of the disc is
described quantitatively by a diffusion equation for the distribution function 
of stars in the space spanned by integrals of motion of the stars, in 
particular the radial action integral and an integral related to the angular 
momentum. Specifically, disc heating by a succession of transient,
`swing amplified' density waves is studied. It is shown that such density
waves lead predominantly to diffusion of stars in radial action space.
The stochastical changes of angular momenta of the stars and the
corresponding stochastic changes of the guiding centre radii of the stellar
orbits induced by this process are much smaller.
\end{abstract}


\section{Introduction}

The shearing sheet (Goldreich \& Lynden--Bell 1965, Julian \& Toomre 1966)
model has been developed as a tool to study the dynamics of galactic discs and
is particularly well suited to describe theoretically the dynamical mechanisms
responsible for the
formation of spiral arms. For the sake of simplicity the model describes only
the dynamics of a patch of a galactic disc. It is assumed to be infinitesimally 
thin and its radial size is assumed to be
much smaller than the disc. Polar coordinates can be
therefore rectified to pseudo-cartesian coordinates and the velocity field of
the differential rotation of the disc can be approximated by a linear shear
flow. These simplifications allow an analytical treatment of the problem,
which helps to clarify the underlying physical processes operating in the disc.

In two previous papers (Fuchs 2001a, b, referred hereafter to 
as papers I and II) I considered the stellardynamical model of a shearing sheet
and discussed the unbounded sheet and then the dynamical consequences, when 
inner boundary conditions are applied. The aim was to give a consistent
theoretical description of pure swing amplification (Toomre 1981) as well as 
exponentially growing modes in the framework of the same model.

It is well known from numerous studies that, if a star disc is perturbed by a
succession of spiral density waves, the stars are scattered randomly by the 
spiral arms and their velocity dispersion grows steadily so that the background
states of the discs are evolving (Julian 1967, Carlberg \& Sellwood 1985, 
Binney \& Lacey 1988, Jenkins \& Binney 1990). Dynamical disc heating has also
been demonstrated in numerical simulations of the dynamical evolution of star 
discs such as by Sellwood \& Carlberg (1984) or Toomre (1990).

In the present paper I discuss disc heating of the shearing sheet. I follow
in particular the theory of diffusion of stars in the two--dimensional action 
integral space due to wave -- star scattering  developed by Dekker (1976). In 
section (2) I briefly describe the formal derivation of the diffusion equation 
in the framework of quasi--linear theory (Hall \& Sturrock 1967) and in section
(3) I calculate diffusion coefficients for scattering of stars by swing
amplified density waves. This kind of spiral density waves has been shown 
to be relevant for disc heating by Toomre (1990), wheras mode--like density 
waves, on the other hand, do not heat effectively (Barbanis \& Woltjer 1967, 
Lynden--Bell \& Kalnajs 1972).

\section{Derivation of the diffusion equation}

When describing the disc heating effects of a succession of transient spiral
density waves, one has to distinguish between the long time scales, on which
the overall distribution function of stars in phase space is evolving,
i.e.~the dynamical heating time scale, from the shorter time scales, on which
the individual density waves develop. As is well known from plasma physics
(Hall \& Sturrock 1967) this concept allows to derive from the Boltzmann
equation in quasi--linear approximation a Fokker--Planck equation for the
long term evolution of the distribution function. I follow here in particular
the adaption of the formalism to the dynamics of stellar discs by Dekker
(1976). The phase space distribution function can be expressed with the aid of
the action and angle variables, $J_{\rm 1}, J_{\rm 2}$ and $w_{\rm 1},
w_{\rm 2}$, respectively,
\begin{equation}
f(J_{\rm 1}, J_{\rm 2}, w_{\rm 1}, w_{\rm 2}; t)\,.
\end{equation}
The evolution of the distribution function with time is determined by the
collisionless Boltzmann equation,
\begin{equation}
\frac{\partial f}{\partial t} + \left[ f , H \right] = 0 \,,
\end{equation}
where $H$ is the Hamiltonian of the stellar orbits, and the square bracket
denotes the usual Poisson bracket. Following Dekker (1976) I define a suitable
mean of the distribution function by averaging it over the angle variables,
\begin{equation}
\langle f \rangle = \frac{1}{4\pi^2} \int^{2 \pi}_0 dw_{\rm 1} \int^{2 \pi}_0 
dw_{\rm 2} f(J_{\rm 1}, J_{\rm 2}, w_{\rm 1}, w_{\rm 2}; t)\,.
\end{equation}
This averaged distribution function will evolve slowly on the long time scale.
For the rapidly fluctuating part of the distribution function $\delta f = f -
\langle f \rangle$ one obtains from the general Boltzmann equation (2)
\begin{equation}
\frac{\partial \delta f}{\partial t} + \frac{\partial \langle f
\rangle }{\partial t} + \left[ \langle f \rangle ,
 H_{\rm 0} \right] + \left[ \delta f , H_{\rm 0} \right]
+ \left[ \langle f \rangle, \delta \Phi \right] + \left[ \delta f , \delta \Phi
\right] = 0 \,,
\end{equation}
where the Hamiltonian has been split up as $H= H_{\rm 0} + \delta \Phi$ with
$\delta \Phi$ denoting the fluctuations of the gravitational potential. The
first Poisson bracket in equation (4) vanishes,
because $\langle f \rangle$ depends only on the
action integrals. The time derivative of the averaged distribution function is
expected to be much smaller than that of the fluctuating part. Thus equation
(4) can be cast into a Boltzmann equation for $\delta f$, and, if the quadratic
term $[ \delta f , \delta \Phi ]$ is neglected, into a linearized Boltzmann
equation,
\begin{equation}
\frac{\partial \delta f}{\partial t} + \left[ \delta f , H_{\rm 0} \right]
= - \left[ \langle f \rangle , \delta \Phi \right]\,,
\end{equation}
which has been widely used to study the dynamics of galactic discs. $\langle
f \rangle$
has taken the role of the axisymmetric, stationary background distribution. Its
slow evolution with time is neglected on short time scales, but taking the
average of the Boltzmann equation (2) leads to
\begin{equation}
\frac{\partial \langle f \rangle}{\partial t} + \langle \left[ \delta f ,
H_{\rm 0}\right] \rangle + \langle \left[ \langle \delta f \rangle ,
\delta \Phi \right] \rangle + \langle \left[ \delta f , \delta \Phi \right]
\rangle = 0\,.
\end{equation}
The second and third terms of equation (6) vanish because $\langle \delta f
\rangle = \langle \delta \Phi \rangle = 0$ by definition,
so that equation (6) takes the form
\begin{equation}
\frac{\partial \langle f \rangle}{\partial t} + \langle \left[ \delta f ,
\delta \Phi \right] \rangle = 0\,,
\end{equation}
which descibes the long--term evolution of the averaged distribution function
in quasi--linear approximation. Equation (7) is also valid, if instead of action
and angle variables other variables are used, as long as $J_1$, $J_2$ are
integrals of motion and $w_1$, $w_2$ their canonical conjugates. 

The potential perturbation $\delta \Phi$ can be Fourier transformed with
respect to the angle variables,
\begin{equation}
\delta \Phi = \int dl_{\rm 1} \int dl_{\rm 2} \delta \Phi_{\rm {\bf l}}({\bf J
};t) e^{i \left( l_{\rm 1}w_{\rm 1} + l_{\rm 2}w_{\rm 2} \right)} \,,
\end{equation}
and similarly
\begin{equation}
\delta f = \int dl_{\rm 1} \int dl_{\rm 2} \delta f_{\rm {\bf l}}({\bf J
};t) e^{i \left( l_{\rm 1}w_{\rm 1} + l_{\rm 2}w_{\rm 2} \right)} \,.
\end{equation}
On the other hand, the lhs of equation (5) represents the total derivative of
$\delta f$ along stellar orbits in the {\em axisymmetric} part of the
gravitational potential. Thus
\begin{eqnarray}
\delta f & = & - \int^t_{t_{\rm 0}} dt' \left[ \langle f \rangle ({\bf J
}_{\rm t'};t') , \delta \Phi({\bf J}_{\rm t'}, {\bf w}_{\rm t'}; t') \right]
\nonumber \\
 & = & \int^t_{t_{\rm 0}} dt' \sum_{\rm n} \frac{\partial \langle f
\rangle }{\partial J_{\rm n}} \Big|_{t'}
 \int dl_{\rm 1} \int dl_{\rm 2} \delta \Phi_{\rm {\bf l}}({\bf J}_{\rm t'}; 
t') \nonumber \\  & & \cdot i l_{\rm n} e^{i \left( l_{\rm 1}w_{\rm 1,t'}
+ l_{\rm 2}w_{\rm 2,t'} \right)} \,,
\end{eqnarray}
where the integration is to be taken along `unperturbed' orbits. The indices of
the action and angle variables indicate that the variables, which are the
independent variables of the distribution function and the gravitational
potential, respectively, must be chosen according the `unperturbed' orbit 
starting at ${\bf J}_{\rm t_0}$, $ {\bf w}_{\rm t_0}$ and terminating at
${\bf J}_{\rm t}$, $ {\bf w}_{\rm t}$. In the next section I will apply
equation (10) to a succession of uncorrelated swing amplification
events of short duration. The typical
integration interval $t - t_{\rm 0}$ will be then much smaller than the
time scale, on which the averaged distribution function $\langle f \rangle$
is evolving and equation (10) can be simplified to
\begin{eqnarray}
\delta f & = & \sum_{\rm n}  \frac{\partial \langle f
\rangle }{\partial J_{\rm n}} \Big|_{t}
\int^t_{t_{\rm 0}} dt'
\int dl_{\rm 1} \int dl_{\rm 2} \delta \Phi_{\rm {\bf l}}({\bf J}_{\rm t},
t') \nonumber \\
 & & \cdot i l_{\rm n} e^{i \left( l_{\rm 1}w_{\rm 1, t'}
+ l_{\rm 2}w_{\rm 2,t'} \right)}\,.
\end{eqnarray}
Comparison of equations (9) and (11) shows that
\begin{eqnarray}
\delta f_{\rm {\bf l}} & = & \sum_{\rm n} \frac{\partial \langle f
\rangle }{\partial J_{\rm n}} i l_{\rm n}
\int^t_{t_{\rm 0}} dt' \delta \Phi_{\rm {\bf l}}({\bf J}_{\rm t}, t') 
\nonumber \\
 & & \cdot e^{i \left( l_{\rm 1}(w_{\rm 1,t'}-w_{\rm 1,t})
+ l_{\rm 2}(w_{\rm 2,t'}-w_{\rm 2,t}) \right)}\,.
\end{eqnarray}

Upon inserting expressions (8) and (9) into equation (7) it is straightforward
to evaluate the Poisson bracket and carry out the averaging with respect to the
angle variables. After some algebra one obtains
\begin{equation}
\frac{\partial \langle f \rangle}{\partial t} = - i \int dl_{\rm 1}
\int dl_{\rm 2} \sum_{\rm n} l_{\rm n} \frac{\partial}{\partial J_{\rm n}}
\langle \delta f_{\rm {\bf l}} \delta \Phi^*_{\rm {\bf l}}\rangle \,,
\end{equation}
where use of the fact has been made that $\delta \Phi_{\rm {\bf -l}} = \delta
\Phi^*_{\rm {\bf l}}$ so that $\delta \Phi$ is real. Since the potential
perturbations are supposed to be a succession of short lived, uncorrelated
fluctuations, it is customary to include into the averaging process an
ensemble average over these fluctuations. Inserting finally equation (12) into
(13) leads to a diffusion equation in action integral space,
\begin{equation}
\frac{\partial \langle f \rangle}{\partial t} = \frac{1}{2} \sum_{\rm m, n}
\frac{\partial}{\partial J_{\rm m}} D_{\rm mn}
\frac{\partial \langle f \rangle}{\partial J_{\rm n}} \,,
\end{equation}
with diffusion coefficients
\begin{eqnarray}
D_{\rm mn} & = & 2  \int dl_{\rm 1} \int dl_{\rm 2} l_{\rm m} l_{\rm n}
\int^t_{t_{\rm 0}} dt' \langle \Phi^*_{\rm {\bf l}}(t) \Phi_{\rm {\bf l}}(t')
\rangle \nonumber \\
 & & \cdot e^{i \left( l_{\rm 1}(w_{\rm 1, t'}-w_{\rm 1,t})
+ l_{\rm 2}(w_{\rm 2,t'}-w_{\rm 2,t}) \right)}\,.
\end{eqnarray}
Equations (14) and (15) are valid for any gravitational perturbations of the
stellar disc with moderate amplitudes and short correlation time scales. 
Very similar relations have been derived by Binney \& Lacey (1988) in a 
different way under more general assumptions. However in Dekker's (1976)
approach the duality of equations (5) and (7) is particularly instructive.

\section{Disc heating by swing amplified spiral density waves}

I apply the formalism of the previous section to calculate the diffusion
coefficients for wave -- star scattering by shearing, swing amplified spiral
density waves. The dynamics of the density waves are modelled by
a shearing sheet made of stars. This describes a patch of a 
thin galactic disc. Its centre orbits the galactic centre at galactocentric 
radius $r_{\rm 0}$ with an angular velocity $\Omega_{\rm 0}$. Pseudo--cartesian 
coordinates are defined with respect to the centre of the patch,
\begin{equation}
x = r - r_{\rm 0},\,y = r_{\rm 0}(\theta - \Omega_{\rm 0} t)\,,
\end{equation}
where $r$ and $\theta$ denote galactic polar coordinates, respectively. 
As explained in paper (I) the
equations of motion of the stars are derived from the Hamiltonian
\begin{equation}
H_{\rm 0} = \frac{1}{2} \dot{r}^2 + \frac{1}{2} r^2_{\rm 0}(\dot{\theta} -
\Omega _{\rm 0})^2 -2 A \Omega_{\rm 0} (r - r_{\rm 0})^2\,,
\end{equation}
or alternatively
\begin{equation}
H_{\rm 0} = \kappa J_{\rm 1} + \frac{A}{2B} J^2_{\rm 2} - \frac{1}{2}
\Omega _{\rm 0}^2 r_{\rm 0}^2\,,
\end{equation}
where $A$ and $B$ denote Oort's constants. The resulting orbits are simple
epicyclic motions, which can be written as
\begin{equation}
x = \frac{J_{\rm 2}}{- 2 B} + \sqrt{\frac{2 J_{\rm 1}}{\kappa}} \sin{w_{\rm
1}},\,y = w_{\rm 2} - \frac{\sqrt{ 2 \kappa J_{\rm 1}}}{2 B} \cos{w_{\rm 1}}\,,
\end{equation}
with $\kappa = \sqrt{ - 4 \Omega_{\rm 0} B}$ the epicyclic frequency. $J_{\rm
1}$ is the radial action integral of an orbit. $J_{\rm 2}$ denotes the
integral $J_2 = \dot{y} + 2 \Omega_0 x$ of an epicyclic orbit and is related to
the angular momentum of a star as $J_2 = (r^2 \dot{\theta} - r_0^2 
\Omega_0)/r_0$. As can be seen from equation (19) the guiding centre radius of
the orbit is given by
\begin{equation}
x_{\rm g} = \frac{J_2}{- 2 B}\,.
\end{equation}
$w_{\rm 1} = \kappa t$ and $w_{\rm 2} = \frac{A}{B}
J_{\rm 2} t$ are variables canonical conjugate to $J_1$ and $J_2$, 
respectively. The radial and circumferential velocities are given by
\begin{equation}
u = \sqrt{2 \kappa J_{\rm 1}} \cos{w_{\rm 1}},\,
v = \frac{2 B}{\kappa} \sqrt{2 \kappa J_{\rm 1}} \sin{w_{\rm 1}}\,,
\end{equation}
where $v$ is defined relative to mean shearing velocity $\dot{y} = - 2 A x$.
The dynamics of a shearing sheet made of stars
has been studied extensively by Julian \& Toomre (1966) (cf.~also Toomre 1981)
and is discussed at length in paper (I) using strictly Eulerian
coordinates. The principal result is that the wave crests of the density waves,
which appear in the disc, swing around following the mean shearing motion of
the stars. While the waves swing around their amplitudes are amplified
transitorily and then die away. In paper (II) it is shown that also mode -- 
like,  quasi--stationary solutions can be constructed for the shearing sheet by
introducing an inner reflecting boundary in the disc. Such kind
of density waves is of no concern in the present context, however, since they
hardly heat the disc at all.

The perturbations of the gravitational potential of the shearing sheet are
customarily Fourier analyzed as
\begin{eqnarray}
&&\delta \Phi = \int dk_{\rm x} \int dk_{\rm y} \delta \Phi_{\rm {\bf k}}
e^{i \left[ k_{\rm x} x + k_{\rm y} y \right]} \nonumber\\ &&
 = \int dk_{\rm x} \int dk_{\rm y} \delta \Phi_{\rm {\bf k}} {\rm exp} i 
\Big[ k_{\rm x} \frac{J_{\rm 2}}{- 2 B} + k_{\rm x} \sqrt{\frac{2 J_{\rm
1}}{\kappa}} \sin{w_{\rm 1}} \nonumber \\ 
 && + k_{\rm y} w_{\rm 2} - k_{\rm y} \frac{\sqrt{2
\kappa J_{\rm 1}}}{2 B} \cos{w_{\rm 1}}\Big]\,.
\end{eqnarray}
The functional dependence on the variable $w_2$ is of the form as in
equation (8) and I use in the following the wave number $l_2$ instead of
$k_{\rm y}$. The dependence on the angle variable $w_1$  
can be easily adapted to the form of equation (8) by taking an inverse
Fourier transform of equation (22) with respect to $w_1$. The resulting Fourier
coefficients are inserted then into equation (15). There is,
however, a difference of the meaning of the variable $w_2$ in this section from
the meaning of the corresponding variable in the previous section. $w_2$
measures now the drift of the guiding centre of the orbit in the 
$y$--direction and
is thus not an angle variable. Averaging the distribution function with respect
to $w_2$ is impractical and I consider in the following the evolution of the
distribution function in ($J_1,J_2$)--space, i.e.~the distribution function
integrated over $w_1$ and $w_2$, respectively, which is conceptually slightly 
different from the average (3). This leads to expressions for the diffusion 
coefficients of the form
\begin{eqnarray}
&&D_{\rm mn} = 2
\int^{2 \pi}_0 dw_1 \int^{2 \pi}_0 dw'_1 \int^\infty_{-\infty} dk_{\rm x}
\int^\infty_{-\infty} dk'_{\rm x}
\int^\infty_{-\infty} dl_{\rm 1} \nonumber \\ && \int^\infty_{-\infty}
dl_{\rm 2}  \int^t_{t_{\rm 0}} dt'
e^{i \big[ -l_{\rm 1} w_{\rm 1}
+ k_{\rm x} \sqrt{\frac{2 J_{\rm 1}}{\kappa}} \sin{w_{\rm 1}}
- l_{\rm 2} \frac{\sqrt{2 \kappa J_{\rm 1}}}{2 B} \cos{w_{\rm 1}}\big]}
\nonumber \\&&
 \cdot l_{\rm m} l_{\rm n} \langle \delta \Phi_{\rm k_{\rm x},
l_{\rm 2}}(t') \delta \Phi^*_{\rm k'_{\rm x}, l_{\rm 2}}(t) \rangle \\ &&
 \cdot e^{- i \big[ l_{\rm 1} \kappa + l_{\rm 2} \frac{A}{B} J_{\rm 2} \big]
(t - t')} e^{i\left(k_{\rm x}- k'_{\rm x}\right) \frac{J_{\rm 2}}{- 2 B}}
\nonumber \\&&
 \cdot e^{- i \big[ -l_{\rm 1} w'_{\rm 1}
+ k'_{\rm x} \sqrt{\frac{2 J_{\rm 1}}{\kappa}} \sin{w'_{\rm 1}}
- l_{\rm 2} \frac{\sqrt{2 \kappa J_{\rm 1}}}{2 B} \cos{w'_{\rm 1}}\big]}
\nonumber \,.
\end{eqnarray}
Note that the diffusion tensor is symmetric by construction. 

In paper (I) it is illustrated how the shearing sheet
responds to internal and external perturbations by developing density waves
with shearing crests and amplitudes amplified while the waves swing around.
Fig.~1 shows the distribution of amplitudes of the potential perturbations as
function of wave numbers $k_{\rm x}, k_{\rm y}$, respectively, calculated using
the formulae of paper (I) (section 11.4) for the case when the swing 
amplification mechanism is fed by white noise. Toomre (1990) discusses the 
source of the noise and argues convincingly that swing amplified white noise 
explains exactly the behaviour of the star discs in his numerical simulations or
that by Sellwood \& Carlberg (1984). The distribution of amplitudes, which 
represents the superposition of many short lived shearing density waves, is
quasi--stationary and can be modelled for positive wave numbers
$k_{\rm y}$ empirically as
\begin{equation}
\delta \Phi_{\rm {\bf k}} = \tilde{\Phi}  e^{- \frac{(k_{\rm x} - k_{\rm
x0})^2}{2 \sigma^2_{\rm k_{\rm x}}} - \frac{(k_{\rm y} - k_{\rm
y0})^2}{2 \sigma^2_{\rm k_{\rm y}}} } \,,
\end{equation}
and continued at negative wave numbers $k_{\rm y}$ as $\delta
\Phi_{\rm {\bf -k}} = \delta \Phi_{\rm {\bf k}}$. The
parameters are estimated as $k_{\rm x0}$ = 1.5 $k_{\rm crit}$, $k_{\rm y0}$ =
0.5 $k_{\rm crit}$, $\sigma_{\rm k_{\rm x}}$ = 0.7 $k_{\rm crit}$, and
$\sigma_{\rm k_{\rm y}}$ = 0.1 $k_{\rm crit}$ for the case $A = -B =
\frac{1}{2} \Omega_{\rm 0}$.
The critical wave number is defined as $k_{\rm crit} =
\kappa^2/(2 \pi G \Sigma_{\rm d})$ with $G$ the constant of gravity and
$\Sigma_{\rm d}$ the surface density of the disc, respectively.
\begin{figure}
\begin{center}
\epsfxsize=8.0cm
   \leavevmode
      \epsffile{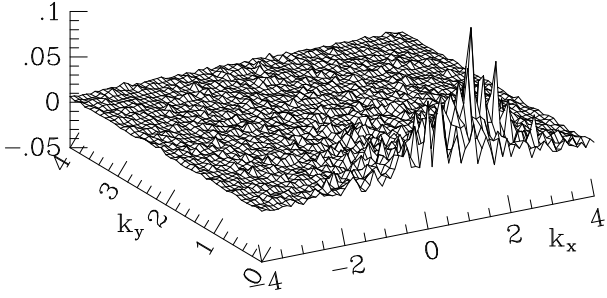}
 \caption{Response of the swing amplifier to continous white noise. The
quasi--stationary distribution of the amplitudes of the potential
perturbations of the star disc, $\sqrt{| \Phi_{\rm {\bf k}} |^2}$, 
is shown as function of radial and
circumferential wave numbers $k_{\rm x}$ and $k_{\rm y}$. At negative wave
numbers $k_{\rm y}$ the distribution of amplitudes continues as $\delta
\Phi_{\rm {\bf -k}} = \delta \Phi_{\rm {\bf k}}$. Each spike represents a
superpostion of
swing amplified density waves all travelling at a constant wave number $k_{\rm
y}$ along the $k_{\rm x}$ abscissae at a speed of $\dot{k}_{\rm x, eff} = 2 A
k_{\rm y}$. The wave numbers
are given in units of $k_{\rm crit}$, whereas the amplitudes are normalized
arbitrarily. The adopted parameters of the disc model are $A = -B = \frac{1}{2}
\Omega_{\rm 0}$ and $Q = 1.4$. See paper (I) for further details.}
         \label{imag}
   \end{center}
   \end{figure}
Using this parametric model the quadratures in equation (23) can be carried out
explicitely. Each spike in Fig.~1 represents the superpostion of
swing amplified density waves all travelling at a constant wave number $k_{\rm
y}$ along the $k_{\rm x}$ abscissae at a speed of $\dot{k}_{\rm x, eff} = 2 A
k_{\rm y}$ with an amplitude which can be modelled as
\begin{eqnarray}
&& \tilde{\Phi}_{\rm k_{\rm x}^{\rm in}, l_2}(t^{\rm in}) e^{-\frac{(k_{\rm 
x}^{\rm in} + 2 A l_2 ( t' - t^{\rm in}) - k_{\rm x_0} )^2}{2 \sigma_{\rm 
k_{\rm x}}^2} -\frac{(l_2 - k_{\rm y 0})^2}{2 \sigma_{\rm k_{\rm y}}^2}} 
\nonumber \\ && \cdot
\delta(k_{\rm x}^{\rm in} + 2 A l_2 ( t' - t^{\rm in}) - k_{\rm x} )\,.
\end{eqnarray}
In equation (25) $k_{\rm x}^{\rm in}$ denotes the original radial wave numbers 
of the waves and $t^{\rm in}$ the
time, when the waves were launched. $\tilde{\Phi}_{\rm 
k_{\rm x}^{\rm in}, l_2}(t^{\rm in})$ parameterizes the excitation rate of the
waves and the peak amplification factor of the swing amplifier (cf.~paper I).
Since white noise is a stationary random process, $\tilde{\Phi}_{\rm 
k_{\rm x}^{\rm in}, k_2}(t^{\rm in})$ is evenly distributed over wave number
space and $t^{\rm in}$. Each of the density waves is only correlated with
itself. This implies, if the autocorrelation function in equation (23)
is considered, restrictions of the wave numbers and the time intervals to
\begin{eqnarray}
&& k_{\rm x}^{\rm in} + 2 A l_2 ( t' - t^{\rm in}) - k_{\rm x}  = 0,\quad 
{\rm and} \nonumber \\ && 
k_{\rm x}^{\rm in} + 2 A l_2 ( t - t^{\rm in}) - k'_{\rm x}
 = 0\quad {\rm so}\, {\rm that} \nonumber \\ && k'_{\rm x} - k_{\rm x} 
- 2 A l_2 ( t - t') = 0\,.
\end{eqnarray}
This constraint and equation (25) lead to an autocorrelation
function in the form of
\begin{eqnarray}
 & &\langle \delta \Phi_{\rm k_{\rm x}, l_2}(t') \delta \Phi_{\rm k'_{\rm x}, 
l_2}(t) \rangle  
\nonumber \\ && = {|\Phi_0|}^2 \delta (k'_{\rm x} - k_{\rm x}  - 2 A l_2 ( t 
- t')) \nonumber \\ && \cdot e^{- \frac{(k'_{\rm x} - k_{\rm x_0})^2}{2 
\sigma^2_{\rm k_{\rm x}}}} e^{- \frac{(k_{\rm x} - k_{\rm x_0})^2}{2 
\sigma^2_{\rm k_{\rm x}}}} e^{ - \frac{(l_{\rm 2} - k_{\rm y_0})^2}
{\sigma^2_{\rm k_{\rm y}}}}\,,
\end{eqnarray}
where $|\Phi_0|^2$ is a normalization constant.
The time integral of the autocorrelation function according to equation (23) is
given by ($l_2 > 0$)
\begin{eqnarray}
 & & \int^t_{t_{0}} dt' \delta (k'_{\rm x} - k_{\rm x}- 2 A l_2 (t -t'))
e^{i[l_1 \kappa  + l_2  \frac{A}{B} J_2] (t' - t)}  \nonumber \\
 & & = \frac{1}{2 A l_2}
e^{-i [ l_1 \kappa + l_2 \frac{A}{B} J_2 ] \frac{k'_{\rm x} - k_{\rm x}}{2 A
l_2} }\,.
\end{eqnarray}
I consider next in equation (23) the integration with respect to wave number
$k_{\rm x}$,
\begin{eqnarray}
&&\int_{-\infty}^{\infty} dk_{\rm x} e^{-\frac{(k_{\rm x} - k_{\rm x_0})^2}{2
\sigma_{\rm k_{\rm x}}^2}} e^{ik_{\rm x}\left[\frac{J_2}{-2B}+
\sqrt{\frac{2J_1}{\kappa}}\sin{w_1}\right]} \nonumber \\ && \cdot
e^{i\left[l_1 \kappa + l_2 \frac{A}{B} J_2\right] \frac{k_{\rm x} - k'_{\rm
x}}{2 A l_2}} \nonumber \\&& = \sqrt{2 \pi} \sigma_{\rm k_{\rm x}}
e^{-\frac{\sigma_{\rm k_{\rm x}}^2}{2} \left[ \sqrt{\frac{2
J_1}{\kappa}}\sin{w_1} + \frac{l_1 \kappa}{2 A l_2}\right]^2} \nonumber \\ &&
\cdot e^{ik_{\rm x_0}\left[\sqrt{\frac{2J_1}{\kappa}}\sin{w_1} + \frac{l_1
\kappa}{2 A l_2}
\right]-i\left[l_1 \kappa + l_2 \frac{A}{B} J_2\right] \frac{k'_{\rm x}} 
{2 A l_2}}\,,
\end{eqnarray}
and similarly
\begin{eqnarray}
&&\int_{-\infty}^{\infty} dk'_{\rm x} e^{-\frac{(k'_{\rm x} - k_{\rm x_0})^2}{2
\sigma_{\rm k_{\rm x}}^2}} e^{-ik'_{\rm x}\left[\frac{J_2}{-2B}+
\sqrt{\frac{2J_1}{\kappa}}\sin{w'_1}\right]} \nonumber \\ && \cdot
e^{-i\left[l_1 \kappa + l_2 \frac{A}{B} J_2\right] \frac{k'_{\rm x}}{2 A l_2}} 
\nonumber \\&& = \sqrt{2 \pi} \sigma_{\rm k_{\rm x}}
e^{-\frac{\sigma_{\rm k_{\rm x}}^2}{2} \left[ \sqrt{\frac{2
J_1}{\kappa}}\sin{w'_1} + \frac{l_1 \kappa}{2 A l_2}\right]^2} \nonumber \\ &&
\cdot e^{-ik_{\rm x_0}\left[\sqrt{\frac{2J_1}{\kappa}}
\sin{w'_1} +\frac{l_1 \kappa}{2 A l_2} \right]}\,.
\end{eqnarray}
In equations (29) and (30) terms of the kind
\begin{equation}
\sigma_{\rm k_{\rm x}}^2 \left(\frac{2 J_{\rm 1}}{\kappa}\right)
\end{equation}
are neglected as quadratically small.
This is justified, because the majority of the stars have epicycle sizes
smaller than the critical wave length $\lambda_{\rm crit}$, the typical spacing
between spiral arms (Julian \& Toomre 1966, cf.~also paper I). The epicycle 
size is determined by $\sqrt{2 J_{\rm 1}/\kappa}$,
whereas $\sigma_{\rm k_{\rm x}} \propto k_{\rm crit} = 2
\pi / \lambda_{\rm crit}$ (cf.~Fig.~1). Similarly terms of the kind 
\begin{equation}
\sigma_{\rm k_{\rm x}}^2\left(\frac{\kappa}{2 A l_2}\right)^2
\end{equation}
will be neglected, because
\begin{equation}
\frac{\sigma_{\rm k_{\rm x}}\kappa}{2 A l_2} \propto \frac{\sigma_{\rm 
k_{\rm x}}}{\dot{k}_{\rm x, eff}} \frac{1}{T_{\rm orb}} \propto \frac{T_{\rm 
acc}}{T_{\rm orb}} \ll 1\,,
\end{equation}
where $T_{\rm acc}$ denotes the duration of disc heating by a
single spike of the potential fluctuations, which is only effective close to the
peak in Fig.~1, whereas $T_{\rm orb}$ is the orbital period of the stars. In
this aspect the disc heating mechanism described here is rather impulsive. Next
I consider the integration in equation (23) with respect to $l_1$, 
\begin{equation}
\int_{-\infty}^{\infty} dl_1 \left\{ \begin{array}{c}
l_1^2\\l_1\\1 \end{array} \right\} e^{-il_1(w_1-w'_1)}\,,
\end{equation}
where the upper row refers to $D_{11}$, the middle row to $D_{12}=D_{21}$, and
the lower row to $D_{22}$, respectively. The results are given by delta 
functions and derivatives thereof,
\begin{equation}
 2 \pi\left\{ \begin{array}{c}
-\delta''(w_1-w'_1)\\i\delta'(w_1-w'_1)\\ \delta(w_1-w'_1) \end{array} \right\}
\,.
\end{equation}
The next step is the integration with respect to the angle variable $w_1$,
which leads for the diffusion coefficient $D_{22}$ to
\begin{eqnarray}
&&\int_0^{2\pi} dw_1 \delta(w_1-w'_1) e^{-i l_2
\frac{\sqrt{2\kappa J_1}}{2B}(\cos{w_1}-\cos{w'_1})} \nonumber \\ &&
 \cdot e^{i k_{\rm x_0}
\sqrt{\frac{2J_1}{\kappa}}(\sin{w_1}-\sin{w'_1})} = 1\,,
\end{eqnarray}
for $D_{12}=D_{21}$ to
\begin{eqnarray}
&& i \int_0^{2\pi} dw_1 \delta'(w_1-w'_1) e^{-i l_2
\frac{\sqrt{2\kappa J_1}}{2B}(\cos{w_1}-\cos{w'_1})} \nonumber \\ &&
\cdot e^{  i k_{\rm x_0}
\sqrt{\frac{2J_1}{\kappa}}(\sin{w_1}-\sin{w'_1})}\nonumber\\ && = - l_2
\frac{\sqrt{2\kappa J_1}}{2B}\sin{w'_1} - k_{\rm x_0}
\sqrt{\frac{2J_1}{\kappa}}\cos{w'_1} \,,
\end{eqnarray}
and for $D_{11}$ to
\begin{eqnarray}
&& - \int_0^{2\pi} dw_1 \delta''(w_1-w'_1) e^{-i l_2
\frac{\sqrt{2\kappa J_1}}{2B}(\cos{w_1}-\cos{w'_1})} \nonumber \\&& 
\cdot e^{ i k_{\rm x_0}
\sqrt{\frac{2J_1}{\kappa}}(\sin{w_1}-\sin{w'_1})}\nonumber\\ && = -
\Big[i l_2 \frac{\sqrt{2\kappa J_1}}{2B}\cos{w'_1} - i k_{\rm x_0}
\sqrt{\frac{2J_1}{\kappa}}\sin{w'_1} \nonumber \\ &&
+\left(i l_2 \frac{\sqrt{2\kappa J_1}}{2B}\sin{w'_1} + i k_{\rm x_0} 
\sqrt{\frac{2J_1}{\kappa}}\cos{w'_1}\right)^2\Big]\,.
\end{eqnarray}
The integration over $w'_1$ gives simply a factor of $2 \pi$ for
$D_{11}$, whereas the integration over the trigonometric functions in equation
(37) leads to the result
\begin{equation}
D_{12} = D_{21} = 0\,.
\end{equation}
Integrating equation (38) with respect to $w'_1$ gives
\begin{equation}
\pi \left( l_2^2 \frac{ 2 \kappa J_1}{4 B^2} + k_{\rm x_0}^2 \frac{2
J_1}{\kappa} \right) \,.
\end{equation}
The final step is the integration with respect to $l_2$,
\begin{eqnarray}
&&\int_{-\infty}^{\infty} dl_2 \frac{1}{l_2} \left\{ \begin{array}{c} 1\\l_2^2 
\end{array} \right\} e^{-\frac{(l_2 -k_{\rm y 0})^2}{\sigma_{\rm k_{\rm y}}^2}}
\nonumber \\ && \cdot
\left\{ \begin{array}{c} \pi\left( l_2^2\frac{2\kappa J_1}{4B^2}+k_{\rm x_0}^2 
\frac{2J_1}{\kappa}\right) \\1 \end{array} \right\} \nonumber \\ && =\left\{ 
\begin{array}{c} \frac{2\pi J_1}{\kappa} k_{\rm y 0} \sqrt{\pi} \sigma_{\rm 
k_{\rm y}}\left( \frac{\kappa^2 }{4B^2}+\frac{k_{\rm x_0}^2 }{k_{\rm y 0}^2}
\right)
\\ k_{\rm y 0} \sqrt{\pi} \sigma_{\rm k_{\rm y}}\end{array} \right\} \,,
\end{eqnarray}
where the upper rows refer to the diffusion coefficient $D_{11}$ and the lower
rows to $D_{22}$, respectively. There is a formal divergence at $l_2$ = 0 in the
second term of $D_{11}$ on the lhs of equation (41). I have chosen to ignore
this and have replaced the integral by a saddle point approximation. The reason
is that the model of disc heating used here (cf.~equation 25) becomes unphysical
for small circumferential wave numbers $l_2$, because the density waves approach
then the WKB limit and become long lived, so that they do not heat the disc 
effectively. Due to the symmetry of the distribution of amplitudes
(24) with respect to ${\bf k}$ the effect of density waves with negative wave
numbers $k_{\rm y}$ can be taken into account by multiplying the diffusion
coefficients by a factor of two. Assembling all results leads a diffusion 
tensor of the form.
\begin{equation}
D_{\rm mn} = D_{\rm 0} \left( \begin{array}{cc}
\frac{1}{\kappa} (\frac{\kappa^2}{4 B^2} + \frac{k^2_{\rm x_0}}{k_{20}^2} ) J_1
& 0\\ 0 & 1 \end{array} \right)\,,
\end{equation}
with
$D_{\rm 0} = 8 \pi^\frac{7}{2} |\Phi_0|^2\sigma^2_{\rm k_{\rm x}}
\sigma_{\rm k_{\rm y}} k_{\rm 20}/A $.

The diffusion equation takes the form
\begin{equation}
\frac{\partial \langle f \rangle}{\partial t} = \frac{1}{2} \frac{\partial}
{\partial J_{\rm 1}} \left( \tilde{D}_{\rm 11} J_{\rm 1} \frac{\partial}
{\partial J_{\rm 1}} \langle f \rangle \right) + 
\frac{1}{2} D_{\rm 22} \frac{\partial^2 \langle f
\rangle}{{\partial J_{\rm 2}}^2}\,,
\end{equation}
where the overhead tilde  means the the $J_1$--dependence of $D_{11}$ has been
written separately.
For this kind of disc heating by transient spiral density waves no
correlation between the diffsion in radial action and angular momentum
space is found. The diffusion equation (43) is highly non--linear, because the
diffusion coefficients depend on the
distribution function $\langle f \rangle$ themselves. In particular,
the effectivity of swing amplification of spiral density waves
depends critically on the value of the Toomre stability parameter $Q$ = $\kappa
\sigma_{\rm u}/(3.36 G \Sigma_{\rm d})$, where $\sigma_{\rm u}$ denotes the
radial velocity dispersion of the stars. Thus, when the disc heats up, the
amplitudes $\Phi_0$ of the density waves and the diffusion coefficients
(42) will decrease and the disc heating rate slows down to zero. Numerical
simulations of the dynamical evolution of galactic discs (Sellwood \& Carlberg
1984, Fuchs \& v.~Linden 1998) have shown that this can happen on comparatively
short time scales, if the discs are left uncooled. Only if the discs are
cooled dynamically by adding stars on low peculiar velocity orbits, the
spiral density wave activity can be maintained at a constant level despite the
rising velocity dispersions. In that case $D_{\rm 0} \approx$ const.~and a
simple solution of the diffusion equation
(43) is found by separation of variables in the form
\begin{eqnarray}
\langle f \rangle & = & \frac{1}{(c_{\rm 1} + \frac{\tilde{D}_{\rm 11}}{2} t)
\sqrt{c_{\rm 2} + \frac{D_{\rm 22}}{2} t}} \nonumber \\
 & \cdot & \exp{- \left\{ \frac{J_{\rm 1}}{c_{\rm 1} + \frac{\tilde{D}_{\rm 11}}
{2} t} + \frac{J^2_{\rm 2}}{c_{\rm 2} + \frac{D_{\rm 22}}{2} t} \right\}}
\end{eqnarray}
with arbitrary constants $c_{\rm 1}$ and $c_{\rm 2}$.

\section{Discussion and Conclusions}

The scattering of stars by shearing, short lived density waves leads to
independent diffusion of stars in radial action -- angular momentum space. This
diffusion process has various implications. The radial action integral is
related to the peculiar velocities of the stars as
\begin{equation}
J_{\rm 1} = \frac{1}{2 \kappa} \left( u^2 + \frac{\kappa ^2}{4 B^2} v^2
\right)\,.
\end{equation}
Thus a distribution function of the form
\begin{equation}
\exp{- J_{\rm 1}/\left( c_{\rm 1} + \frac{\tilde{D}_{\rm 11}}{2} t \right)}
\end{equation}
implies a $\exp{- u^2 / 2 \sigma^2_{\rm u}(t)}$ dependence of the velocity
distribution with a predicted radial velocity dispersion of $\sigma_{\rm u}(t)$ 
= $\sqrt{\kappa ( c_{\rm 1} + \tilde{D}_{\rm 11} t /2)}$. Such a rise
of the velocity dispersion with time fits ideally to the actual age--velocity
dispersion relation observed in the solar neighbourhood (cf.~Fuchs et al.~2001
for a recent review of the observational data). Unfortunately the diffusion
coefficients cannot be estimated quantitatively, because the constant
$|\Phi_0|^2$, which parameterizes the white noise, is not known a priori.
But judging from the shape of the heating law (46) wave -- star
scattering of the kind discussed here might well have played an important
role in the Milky Way disc. Whether this is the only disc heating mechanism is
still a matter of debate (Fuchs et al.~2001).

The diffusion of the guiding centre radii of the stellar orbits can be
estimated from the ratio of the diffusion coefficients,
\begin{equation}
\frac{D_{\rm 22}}{D_{\rm 11}} = \frac{\kappa}{J_{\rm 1}} \frac{1}
{\frac{\kappa^2}{4 B^2}+\frac{k^2_{\rm x_0}}{k^2_{20}}}\,.
\end{equation}
As shown above $\kappa D_{\rm 11} t/(2 J_{\rm 1})$ is proportional to 
the square of the radial velocity dispersion, $\sigma^2_{\rm u}(t)$. According 
to epicyclic theory (cf.~equation 20)
\begin{equation}
\langle x^2_{\rm g} \rangle = \frac{1}{4 B^2} \langle J^2_{\rm 2} \rangle \,,
\end{equation}
so that the dispersion of the guiding centre radii of the stellar orbits is 
given by
\begin{eqnarray}
&&\sqrt{\langle x^2_{\rm g} \rangle}=\frac{1}{- 2 B} \sqrt{\frac{D_{\rm
22}}{4} t} 
=\frac{1}{- 2 B} \frac{\sigma_{\rm u}}
{\sqrt{2\left(\frac{\kappa^2}{4 B^2}+\frac{k^2_{\rm x_0}}{k^2_{20}}\right)}}
  \,.
\end{eqnarray}
Using the parameter values estimated above equation (49) implies
$\sqrt{\langle x^2_{\rm g} \rangle} = 0.2 \sigma_{\rm u}/\Omega_{\rm
0}$, if a flat rotation curve is assumed. A star with an age of 5 Gyrs like
the Sun has typically in the solar neighbourhood a velocity dispersion of 50
km/s. Thus $\sqrt{\langle x^2_{\rm g} \rangle}$ = 400 pc = 0.05 $r_{\rm
0}$ in the solar neighbourhood, if a local angular velocity of
$\Omega_{\rm 0}$ = 26 km/s/kpc is adopted. This confirms the conclusion of
Binney \& Lacey (1988) that the diffusion of guiding centre radii driven by a
rapid succession of spiral density waves is rather small. This
assumes that disc heating by transient spiral density waves is the
only disc heating mechanism. However, if the Sun has indeed drifted from its
birth place nearly 2 kpc radially outwards to its present galactocentric radius
as suggested by Wielen, Fuchs, \& Dettbarn (1996) and Wielen \& Wilson (1997),
this would mean that there must be other dynamical heating mechanisms of the
galactic disc. It was shown by Fuchs, Dettbarn, \& Wielen (1994)
theoretically and by numerical simulations that Spitzer--Schwarzschild
diffusion of stars due to gravitational encounters with massive molecular
clouds, for instance, leads to a much more pronounced diffusion of the guiding 
centre radii of the stellar orbits, even if the mechanism does not heat the 
disc effectively.

\section*{Acknowledgments}

I thank A.~Just and R.~Wielen for helpful discussions. I am also grateful to the
anonymous referee, whose comments lead to an improvement of the paper.

\bsp

\label{lastpage}

\end{document}